\documentclass[final,5p,times,twocolumn]{elsarticle}

\usepackage{amssymb}
\usepackage{lineno}
\usepackage{xcolor} 
\usepackage{soul}

\journal{Acta Astronautica}

\begin{document}

\begin{frontmatter}



\title{On Optimal Geometry for Space Interferometers}


\author[inst]{A.~G.~Rudnitskiy\corref{cor}}
\author[inst]{M.~A.~Shchurov} 
\author[inst]{S.~V.~Chernov}
\author[inst]{T.~A.~Syachina}
\author[inst]{P.~R.~Zapevalin}

\affiliation[inst]{organization={Astro Space Center, Lebedev Physical Institute, Russian Academy of Sciences},
            addressline={Profsoyuznaya str. 84/32}, 
            city={Moscow},
            postcode={117997}, 
            country={Russian Federation}}

\cortext[cor]{Corresponding author: almax1024@gmail.com}

\begin{abstract}
This paper examines options for orbit configurations for a space interferometer. In contrast to previously presented concepts for space very long baseline interferometry, we propose a combination of regular and retrograde near-Earth circular orbits in order to achieve a faster filling of $(u,v)$~coverage. With the rapid relative motion of the telescopes, it will be possible to quickly obtain high quality images of supermassive black holes. As a result of such an approach, it will be possible for the first time to conduct high quality studies of the supermassive black hole close surroundings in dynamics.
\end{abstract}
 


\end{frontmatter}


\section{Introduction}
\label{sec:intro}
During the past couple of decades, the very long baseline interferometry (VLBI) method has made significant progress in terms of its capabilities \cite{Gurvits2020}. Here is an example of the global VLBI reaching beyond the Earth with the successful launch of the VSOP \cite{Lovell1999}. Further, the Radioastron mission provided an unprecedented angular resolution of 8~$\mu$as \cite{Kardashev2013}. New concepts for similar observatories were also developed, such as the ARISE space-VLBI mission and VSOP-2 \cite{Ulvestad2000,Mochizuki2003}.

The transition to higher frequencies led to the development of the Event Horizon Telescope (EHT) -- a new global VLBI telescope operating at millimeter and sub-millimeter frequencies. First time in astronomy, it was possible to obtain images of a bright photon ring in Sgr~A* and M87* with a resolution of about 25~$\mu$as \cite{EHT2019a,EHT2022a}. In the coming years, it is planned to develop the Event Horizon Telescope into a next generation (ngEHT) network with increased telescope numbers and operating frequencies (up to 345 GHz) \cite{Palumbo2018,Raymond2021}. The VLBI experiments have advanced by several orders of magnitude since the first ground observations, resulting in increased bandwidth, sensitivity, and quality of the data.
 
In the last few years, conceptual studies of space-VLBI at frequencies of the EHT and ALMA have become the most relevant, since ground VLBI has some insurmountable limitations when it comes to the maximum angular resolution and coherent integration time necessary for high frequency observations to yield a satisfactory signal-to-noise ratio. As a point of reference, it is worth to mention the Millimetron space observatory, which is currently being developed. The observatory will carry a deployable, active cooled, 10-meter antenna capable of supporting space-ground VLBI observations at frequencies up to 345 ~GHz in the Lagrange point L2 of the Sun-Earth system \cite{Novikov2021}.

Due to limitations associated with the Earth's atmosphere for millimeter and sub-millimeter wavelengths, proposals for a purely space-based interferometer -- space-space VLBI -- are being actively developed. These systems could provide fundamentally new information about the closest SMBHs, Sgr~A* and M87* and other sources. Detailed information regarding the proposed concepts for space interferometers is provided below.

Throughout this work, we examine the issues associated with choosing the optimal geometry and propose our vision for orbital configuration for space-space interferometers to perform effective imaging of SMBHs. It is important to note that we do not discuss issues relating to the overall system design and only address those aspects that have an impact on the choice of orbit.

\subsection{Space Interferometer Concepts}
\label{sec:concept}
Space-space VLBI involves the use of several radio telescopes in different orbits simultaneously. The advantages of this approach are obvious: the ability to obtain angular resolutions unattainable by ground methods; without an atmosphere, the frequency threshold above 345 GHz can be crossed and coherent integration times can be achieved even at terahertz frequencies. Further, high frequency observations minimize scattering effects, which are noticeable for Sgr~A* even at  230-345~GHz \cite{Roelofs2019}.

Within the last decade, a variety of proposals and concepts have emerged regarding the construction and launch of space VLBI instruments. One of the modern concepts to be proposed is the Space Millimeter VLBI Array (SMVA), which operated at 8, 22 and 43 GHz at the beginning of its operation \cite{Hong2014}. The maximum angular resolution would be 20~$\mu$as at 43~GHz. This concept will have three development stages, ultimately resulting in a system of three to four 12-15 meter antennas operating at millimeter and sub-millimeter wavelengths (from 345~GHz and above). As described in the concept, it uses elliptical near-Earth orbits that provide relatively sparse $(u,v)$ coverage with a one-day accumulation time.

The success associated with the observation of SMBH in M87* and Sgr~A* by the EHT, formulated a proposal for the further development of the EHT in the form of a space system - Event Horizon Imager (EHI) \cite{Roelofs2019,Fish2020,Kudriashov2021a,Kudriashov2021b}. As part of this concept, two or three space telescopes will be located in polar or equatorial circular medium Earth orbits (MEOs) operating at 230, 345 and 690 GHz, which are compatible with ground ngEHT. EHI will provide the best coverage for $(u,v)$, however its accumulation time is approximately one month.

A more recent proposal is TeraHertz Exploration and Zooming-in for Astrophysics (THEZA)
\cite{Gurvits2021,Gurvits2022}. Interferometers of this type are intended to have scalable configurations and cover frequencies up to approximately 1~THz. 
In this concept, one of the most distinctive features is the compact phased antenna array located on the spacecraft/space platform. 
The paper, however, does not give specific solutions to the geometry of such interferometers, mentioning only low and medium Earth orbits (LEO\& MEO) and possible assembly and launch from orbital stations.

One of the recent concepts is the Event Horizon Explorer, currently under study. It is assumed that this will be an extension of the EHT by adding a space-based node. \cite{Kurczynski2022}.

Furthermore, another recent concept of a space-only high frequency VLBI network is CAPELLA, which can be found in \cite{Trippe2023}. According to the proposal, four radio telescopes should be located at low polar near-Earth orbits orthogonal to each other. The observing frequency is proposed to be only 690~GHz to operate exclusively in space without ground antennas.

Most of concepts share similar characteristics. Typically, two to four spacecraft are required and the interferometer geometry is constructed on near-Earth orbits, namely low and medium circular near-Earth orbits \cite{Roelofs2019,Fish2020,Kudriashov2021a,Kudriashov2021b,Trippe2023} and elliptical orbits \cite{Hong2014}. Despite the fact that we are discussing pure space VLBI, any such system, in principle, can participate in joint observations with ground-based telescopes, which does not fundamentally complicate the system. If such observations are feasible in terms of ground antenna compatibility and limitations.

It has its own challenges, however, when it comes to pure space VLBI. The dimensions of the rocket fairings limit the size of a single antenna. Multiple launch vehicles may be required to launch several large antennas. For a large deployable aperture system to achieve adequate aperture efficiency at higher frequencies, a complex system of adaptation and control of the antenna surface will be required. In addition to telescope lifetimes, which are much shorter than those of ground antennas, there is the issue of data transmission to the Earth or processing on board. Taking a balanced approach to such a project is essential, since all of these factors can significantly increase its cost and complexity.

\subsection{Scientific Tasks}
One of the most important tasks of VLBI are associated with SMBH. The main objective is to obtain detailed images with a much higher angular resolution and frequency than the EHT, while maintaining a high rate of observations repeatability.

The primary target sources could be Sgr~A* and M87*. Even so, the space interferometer science should not be limited to a one or two sources. Using a space interferometer increases angular resolution and solves the scattering problem by switching to a higher frequency. Consequently, space interferometry can provide high fidelity imaging of SMBHs in this context.

With high repeatability of observations, another task related to SMBH is possible only with a space interferometer. In this case, it involves dynamic imaging, namely serial snapshot observations of the SMBH \cite{Andrianov2021,Johnson2023,Ricarte2023,Emami2023}. Our study considers Sgr~A* for dynamic imaging, which is a highly variable source. To achieve an acceptable snapshot image fidelity, either the $(u,v)$ must be densely covered right away or the $(u,v)$ must be filled rapidly. The latter could be achieved through the use of space VLBI.

Space interferometry science cases mentioned above require frequencies higher than those available on the ground, i.e. above 345~GHz. The 230 GHz and 345 GHz bands may be included in order to be compatible with ground antennas. Geometry should provide a better angular resolution than EHT. Several simulations of the Millimetron space-ground interferometer have shown that space-ground baselines can already achieve a resolution about five times better than the EHT \cite{Andrianov2021,Likhachev2022}. Thus, a pure space interferometer should be able to achieve even better angular resolution and observe exclusively at frequencies that are not accessible from the ground. Nevertheless, it is important to consider whether such parameters are necessary. Additionally, higher-resolution observations of other SMBHs may be possible, since their shadows are smaller than those of M87* and Sgr~A* ($\sim$42 and $\sim$53~$\mu$as, respectively). Even direct imaging of binary SMBH systems would be possible.

However, this work first of all, addresses the problem of finding orbital configurations that would allow relatively quickly to obtain a good $(u,v)$~coverage for the synthesis of high-quality SMBH images. As an example to demonstrate the imaging capabilities of the proposed interferometer configuration, we have chosen Sgr~A*.

\section{Interferometer Geometry}
\label{sec:geomtery}
Firstly, let us examine the general concept of the future space interferometer and some of its parameters that will be relevant to this study. It primarily intersects with previously proposed concepts regarding instrumentation, especially those proposed by \cite{Trippe2023}. The operating frequencies could be 230, 345, 690~GHz or even higher. It is possible to have one or more operating bands with simultaneous multi-frequency synthesis \cite{Han2012, Han2017, Rioja2023}.

To perform phase closure, which is essential for VLBI imaging observations, at least three spacecraft should be deployed; however, it is preferable to have four satellites in order to perform amplitude closure as well. Satellites should be equipped with antennas measuring approximately 3-4 meters in diameter.

With such a diameter, telescopes can be placed directly under the fairing without using deployable antennas. By packing a pair of such space telescopes under the fairing, orbital insertion costs could be reduced. Speaking about antenna performance, the surface quality of 3-4 meter solid parabolic antennas reached $\approx$5~$\mu$m~root mean square (RMS), which meets the requirements for successful high frequency operation \cite{Golubev2020}. 

\subsection{Configuration}
\label{sec:orbits}
We propose a combination of regular and retrograde near-Earth circular orbits for two, three, and four space radio telescopes. 

The main idea is to have a relatively high rate of fill for $(u,v)$~coverage without significant loss of quality. This rate of coverage filling could be achieved by the space telescopes moving along retrograde and regular orbits at high relative velocities. 

\begin{figure*}[ht]
    \centering
    \includegraphics[width=0.38\linewidth]{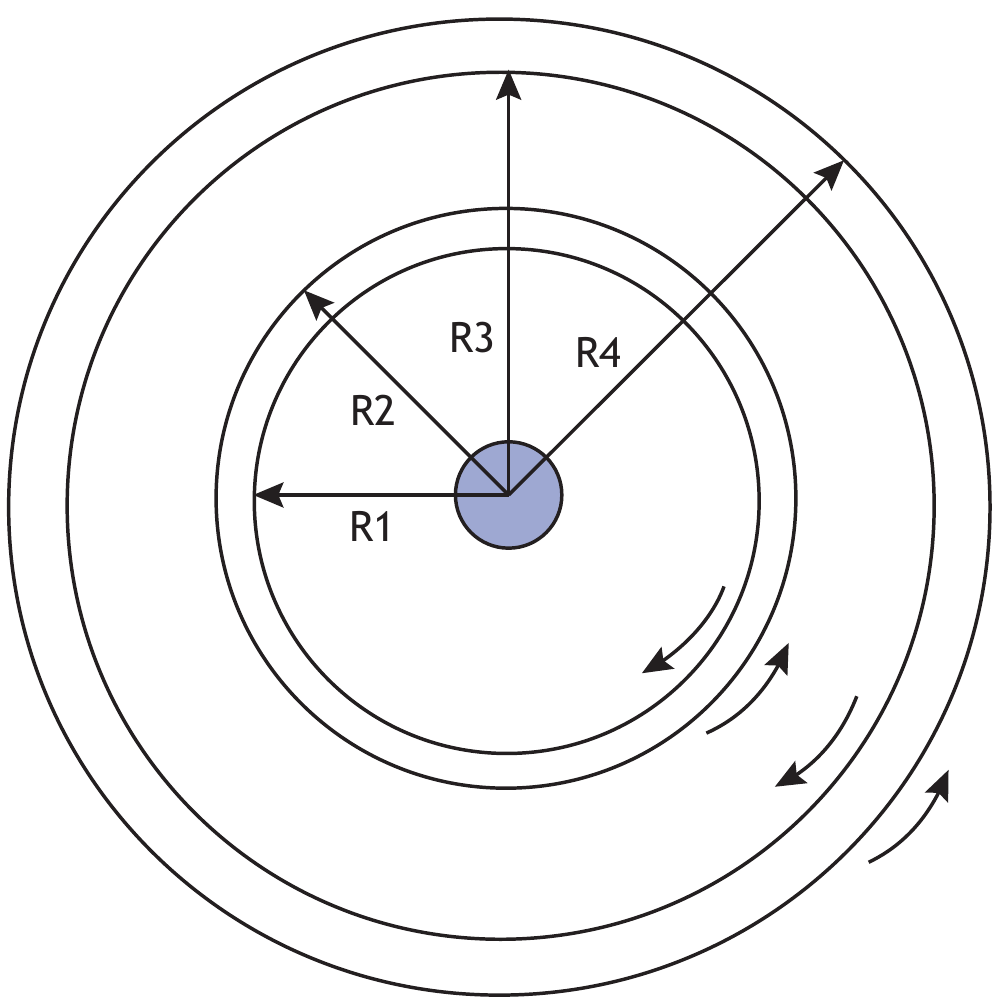}
    \includegraphics[width=0.49\linewidth]{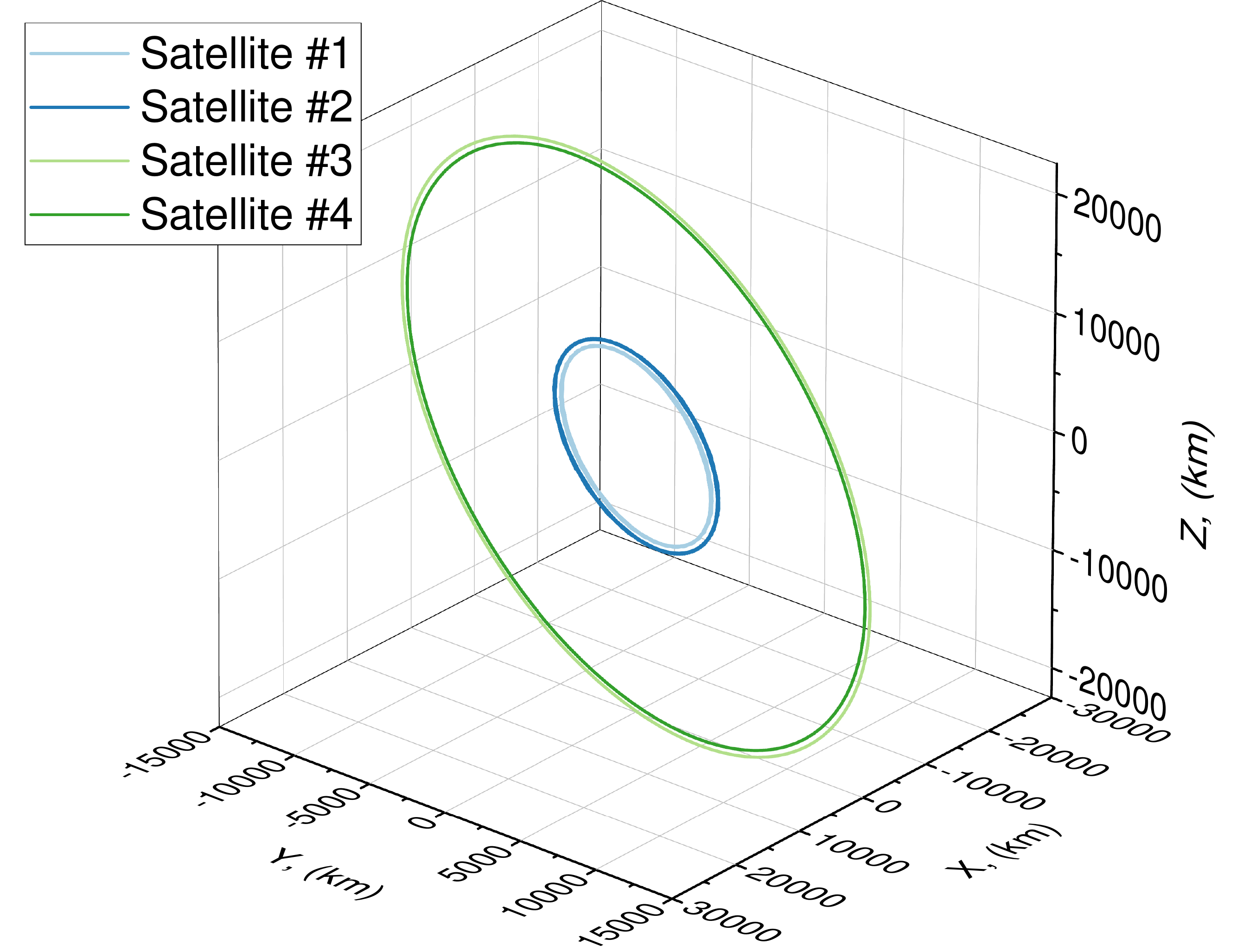}
    \caption{(left): Scheme of space interferometer configuration with four satellites. Radii of orbits: $R_{1}=7500$~km, $R_{2}=8000$~km, $R_{3}=22500$~km, $R_{4}=23000$~km. Arrows show relative direction of satellite motion. Clockwise arrows correspond to regular orbits, counterclockwise arrows correspond to retrograde orbits. (right) Three dimensional representation of integrated orbits (1~period) for 4~satellites in J2000~coordinate system.}
    \label{fig:orbit}
\end{figure*}

In the case of two satellites, two circular orbits of $R1=7500$ and $R2=8000$~km are proposed, one of which is retrograde. It is proposed to add one more regular orbit of 22500 km for three satellites, and one more retrograde orbit of 23000 km for four satellites. Fig.~\ref{fig:orbit} (left) shows the scheme of orbit configuration for four space radio telescopes. Fig.~\ref{fig:orbit} (right) shows the three-dimensional representation of the integrated orbits (one period) for four space radio telescopes. The plane of all orbits is oriented by the normal to the chosen target source Sgr~A*. 

Inner orbits ($R_{1}=7500$ and $R_{2}=8000$ km), regular and retrograde, provide baseline projections ($B$), proportional to $B_{min}^{in}\sim|R_{1} - R_{2}|$, $B_{max}^{in}\sim |R_{1}+R_{2}|$. As a result, the baseline projection changes from a minimum to a maximum value for $\sim$1/4 of the inner orbit period. For the next $\sim$1/4 period these two satellites get closer to each other, thus the baseline projection gets smaller. Because the orbital radii are a little different for these two satellites ($\sim500~$km), the baseline projection rotates on the $(u,v)$~plane every $\sim$1/2 period. In the case of equal radii there would be a constant straight line. However, in the considered case there is a curve passing at some distance from the center of the $(u,v)$~plane, depending on the difference in the radii of the orbits of the two satellites. The larger distance between the two inner orbits could provide faster $(u,v)$coverage evolution, but with larger gaps. Furthermore, the smaller distance provides a better coverage (by angle), but slower evolution. Thus, the distance between regular and retrograde orbits of $\sim$500~km was found to be the best. The parameters proposed orbits are shown in Table~\ref{tab:prec}.

\begin{table}[ht]
    \centering
    \begin{tabular}{|c|c|c|c|}
    \hline
    \begin{tabular}[c]{@{}c@{}}Orbit radius,\\ (km)\end{tabular} & \begin{tabular}[c]{@{}c@{}}Inclination,\\ (deg)\end{tabular} & \begin{tabular}[c]{@{}c@{}}Orbit\\ period,\\ (hrs)\end{tabular} & \begin{tabular}[c]{@{}c@{}}Precession\\ period,\\ (days)\end{tabular} \\ \hline
7500                                                        & -61                                                         & $\sim$1.8                                                          & $\sim$120                                                            \\ \hline
8000                                                        & 119                                                         & $\sim$2.0                                                          & $\sim$150                                                            \\ \hline
22500                                                       & -61                                                         & $\sim$9.3                                                          & $\sim$5621                                                           \\ \hline
23000                                                       & 119                                                         & $\sim$9.6                                                          & $\sim$6071                                                           \\ \hline
    \end{tabular}
    \caption{\label{tab:prec}Parameters of proposed orbits. RAAN = -3.6$^{\circ}$}
\end{table}

Adding the third satellite allows for a better $(u,v)$~coverage and increase in angular resolution. Putting a satellite in an orbit with the same parameters as an inner regular orbit but with a larger radius will result in longer baseline projections. It is therefore necessary to determine the correct radius of the third satellite in order to obtain an optimal $(u,v)$~coverage. If that radius $R$ is three times larger than the inner orbit radius, this will yield additional baseline projections in a range from $B_{min}^{out}$~=~$B_{max}^{in}$ because of the distance between these satellites is $2R$ at their closest point of inner and outer orbits. The maximum baseline B$_{max}^{out}$ in this case will be proportional to the sum of inner and outer radii B$_{max}^{out}\sim~|R_{4}~+~R_{1}|\approx4R_{1}$.

The fourth satellite, which is needed for amplitude closure, could be added to the outer retrograde orbit. Again, in this case $(u,v)$~coverage will be improved as well as angular resolution. Besides satellites $3$ and $4$ form a new broader set of rays, similar to a system of the first pair of satellites. This is due to the oncoming movement of these satellites in outer orbits.

These specific orbits were integrated numerically in a complete force model for 4000 days ($>$10~years), that includes: central and non-central field of Earth, EGM96 model up to 33 harmonics, influence of the Solar system bodies (DE430 ephemeris: all planets, the Sun, the Moon and Pluto) \cite{Folkner2014,EGM96}. The numerical integration was performed using the Runge-Kutta method of the 7(8)-th order with a variable adaptive step.

Our calculations show that the orbits have been stable for over 10 years (with the exception of RAAN, which is influenced by precession). Orbit precession period is shown in Table \ref{tab:prec} (rightmost column). It is worth noting that the precession period is much longer than the orbit period ($\sim$1500 times for inner orbits, and $15000$ times for outer orbits), so we can conclude that the orbits are stable for observing single sources without significant changes in the $(u,v)$-coverage. Observability of a single target source will last $\sim$30 days. This is because the orbit plane will be perpendicular to the source direction twice during the precession period. Due to precession, the normal vector of the orbit plane moves along the RAAN axis, making it possible to observe different target sources.

In considered cases (2, 3 and 4 telescopes), one can take virtually any convenient time when all 4 orbits are coplanar and their normal are directed towards the target source within a range of $\pm$30 degrees. For R1 orbit, the Earth overlaps the source at an angle of about 31.7$^{\circ}$. In the case of two or three satellites, all baseline projections will be obtained within ~20 hours from any moment of launch. Configuration with four telescopes has observation period 30 times longer than outer orbits (R3 and R4). That fact completely eliminates the need to choose a specific start time and synchronize telescope orbits.

\begin{figure*}[ht]
    \centering
    \includegraphics[width=1\linewidth]{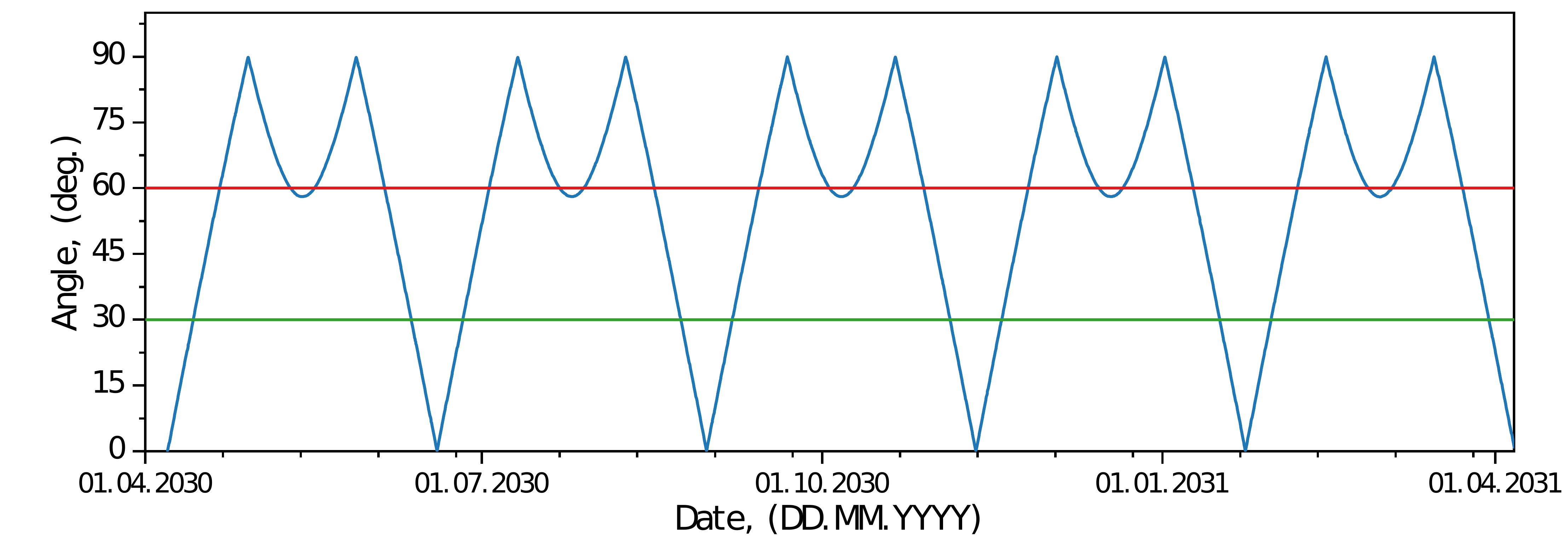}
    \caption{Angle between normal vectors for R1 and R2 orbits in 1 year. Green line shows an acceptable angle limit. Red line shows the upper limit for VLBI observations.}
    \label{fig:coplan}
\end{figure*}

The co-planarity of these orbits should be discussed separately. Because the precession period of R3 and R4 is more than 15 years, co-planarity is relevant only for R1 and R2. Fig.~\ref{fig:coplan} shows the evolution of angle $\beta$ between normal of R1 and R2 orbits during one year. As it can be seen, orbits R1 and R2 will be co-planar approximately every 72 days. As the orbits precess, other sources can be observed. It is true when the orbital planes are co-planar or close to co-planar within $\beta=\pm30^{\circ}$ (green line in Fig.~\ref{fig:coplan}) and the direction to the source. Hence, the $(u,v)$ coverage won't be affected significantly leading to slight asymmetry and shift of the $(u,v)$ points by a factor of $\sqrt{3}/2$. Increase in $\beta$ up to $\pm$ 60$^{\circ}$ (red line in Fig.~\ref{fig:coplan}, 30$^{\circ}$ for each orbit in opposite directions) will lead to a significant decrease in the symmetry of $(u,v)$ coverage (up to 2 times) while still leaving the possibility to conduct observations. The frequency of observations of a particular source will depend on the requirements for symmetry of the $(u,v)$ coverage. Of course, it is possible to perform synchronizing orbital corrections. However, this is a separate study that is beyond the scope of this work.

Thus, taking into account constraints related to the Earth overlapping for R1 and evolution of RAAN only during precession, there will be additional constraints that will allow observation of sources only with a declination within $\pm$30$^{\circ}$.

Retrograde orbits have already been used by Earth remote sensing satellites in the past (e.g. Sentinel-2, Landsat-8; see \cite{Ippolito2008,Warner2009,Cakaj2009,Li2017}). This approach allows for more frequent observations of the same areas. Despite this, most satellites, both commercial and scientific, use regular orbits to reduce fuel costs during launch. For instance, when launching from the equator to a regular or retrograde orbit, the difference in $\Delta v$ is approximately 1 km/s between these orbits. However, such fuel costs are bearable, and it is feasible to launch space telescopes into such retrograde orbits. 

\subsection{Orbit Determination \& Visibility}
Radio interferometry has its own data processing peculiarities. VLBI baseband data are first correlated. In the case of space telescopes, there are certain requirements for the accuracy of the orbit determination for the data processing to be successful. They arise from searching the correlation fringe between the space telescopes. To find a fringe, it is necessary to know exactly the radio signal delay between the radio telescopes. In case of space interferometer, this depends on the orbit determination accuracy. In the correlator, the delay is tracked with the sampling of data. The difference between the true and the model delay $\Delta \tau$ for the integration time $T_{int}$ must not exceed the duration of one or several samples. The period of one sample is equal $1/2\Delta f$, where $\Delta f$ is channel bandwidth. If this difference is constant over the integration interval, it indicates an error in the interferometer baseline determination, i.e. the coordinates of the radio telescopes. If the fringe regularly drifts along the integration time, it means that there is an inaccuracy in the baseline change rate, i.e. the spacecraft's velocity. Equations for the maximum allowed error of spacecraft position and velocity are provided in \cite{Andrianov2014, Zhamkov2016}.

To increase fringe detection probability, consider the number of channels $N_{ch}$ along the delay and fringe rate. The delay corresponding to the fractional part of sample time interval is corrected after Fourier transform in the spectral domain. Next, the phase of signal is shifted to stop interference fringes. This process is called fringe rotation \cite{Likhachev2017}.

The number of Fourier transform channels in the correlator determines the length of the processed data interval in seconds: $T=2N_{ch}/2\Delta f=1/f_0$, where $f_0$ is the operational frequency. Considering the delay $\Delta \tau$ as a Taylor series, and limiting it to first-order terms, the following constraints can be derived for the accuracy of determining the baseline $\Delta B$ and its rate of change $\Delta B/dt$:
\begin{equation}
 \Delta B = \Delta \tau\cdot c<\frac{N_{ch}\cdot c}{2\Delta f},\quad 
 \frac{\Delta B}{dt}<\frac{N_{ch} \cdot c}{2T_{int}f_0}.   
\end{equation}

Thus, the spacecraft estimated orbit position $\Delta B$ and velocity $\Delta B/dt$ errors are shown in the Table \ref{tab:OD}. This orbit determination accuracy could be ensured by global navigation satellite system (GNSS) measurements \cite{MEO2, MEO3, MEO4} and/or VLBI tracking \cite{Duev2012, Klopoter2020}. Based on these values, it is clear that for such orbits it is quite possible to position the spacecraft with the required accuracy. This is especially in the case of ground data processing.

\begin{table}[ht]
    
    \centering
    \begin{tabular}{|c|c|c|c|c|}
    \hline
    \begin{tabular}[c]{@{}c@{}}$N_{ch}$\end{tabular} & \begin{tabular}[c]{@{}c@{}}$\Delta B$ (m)\end{tabular} 
      & \begin{tabular}[c]{@{}c@{}}$\Delta B /dt$ (mm/s) \end{tabular}\\ \hline
32     & 1.2    & 6                                                             \\ \hline
128                                                       & 4.8    & 28                                                                      \\ \hline
256                                                      & 9.6     & 56                                                                      \\ \hline
1024                                                      & 38.4  &222                                                                         \\ \hline
2048                                                      & 76.7   & 445                                                                        \\ \hline
4096                                                      & 153.5  & 890                                                                        \\ \hline
    \end{tabular}
    \caption{\label{tab:OD}Estimated spacecraft position $\Delta B$ and velocity $\Delta B/dt$ errors. $T_{int}$~=~100~sec. Bandwidth~=~4~GHz. Operational frequency~=~690~GHz.}
\end{table}

\begin{figure*}[ht]
    \centering
    \includegraphics[width=0.49\linewidth]{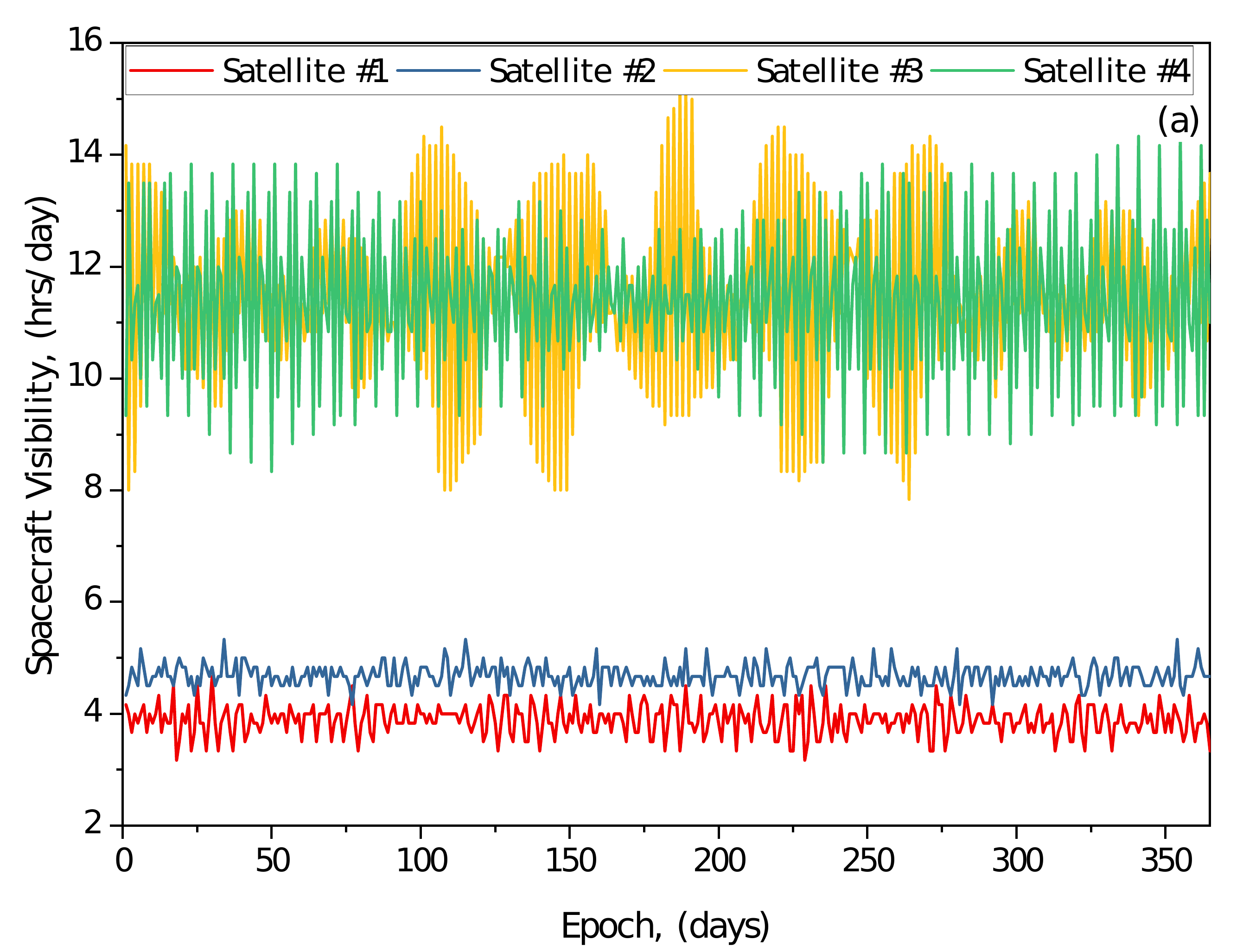}
    \includegraphics[width=0.49\linewidth]{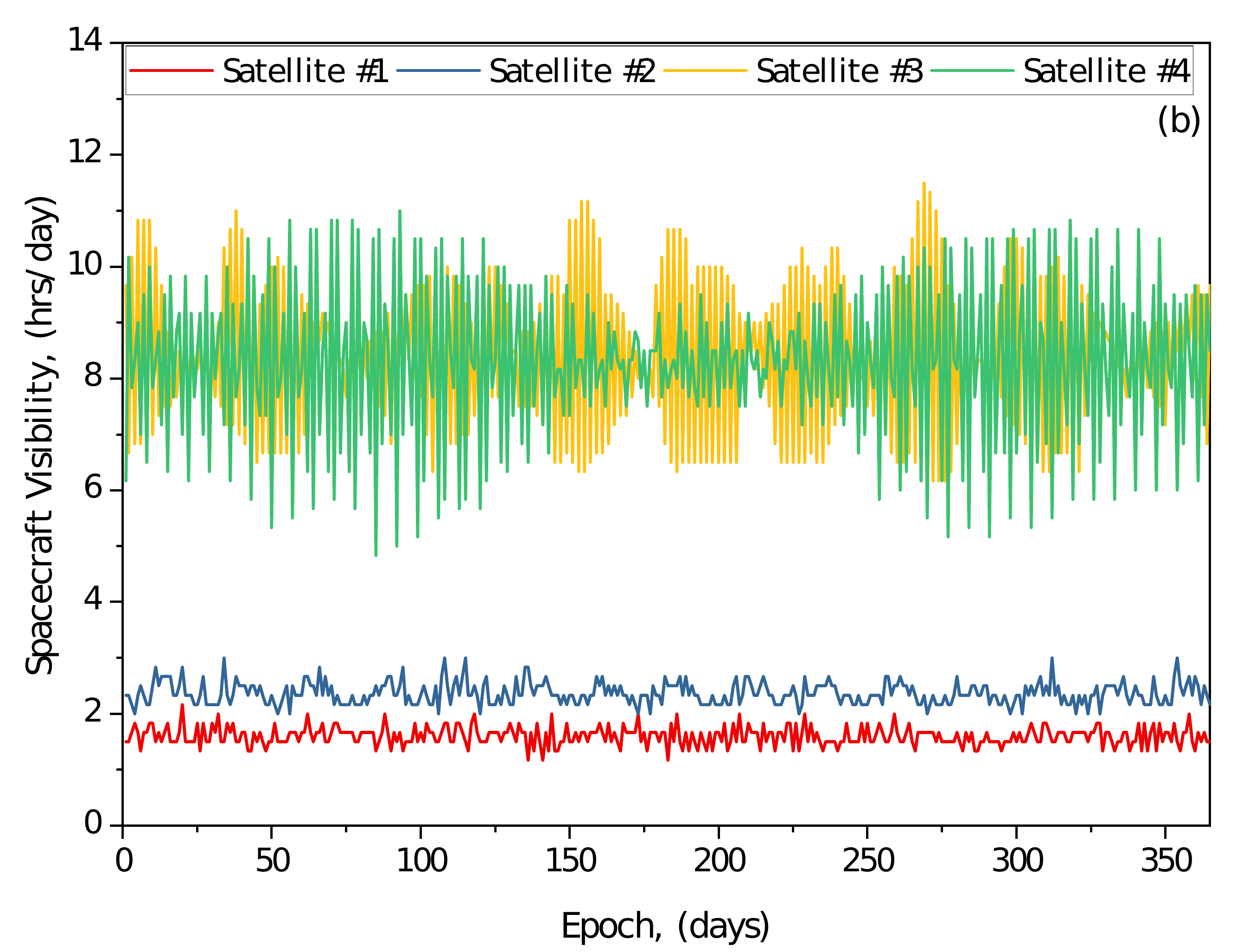}
    \caption{Duration of satellite radio visibility from the Earth tracking stations: (a) -- visible by at least one tracking station, (b) -- visible by at least two stations simultaneously. Four approximate locations were chosen for tracking stations: Eastern Europe (Russia, Bear Lakes), Asia (Kazhahstan, Baikonur), Northern Caucasus (Russia, Zelenchyuk) and Siberia (Russia, Ussuriisk).}
    \label{fig:vis}
\end{figure*}

A measurement of the spacecraft visibility from ground tracking stations has been made. The satellite visibility was calculated over 1 year. Satellites were considered observable and visible by a telescope if they had an elevation of more than $7^{\circ}$ above the horizon. A topocentric coordinate system was introduced using the WGS-84 geodetic system. Fig.~\ref{fig:vis} shows daily visibility in hours per day of satellites by at least one (a) and at least two (b) out of four tracking stations located in the Northern hemisphere for one year. Four approximate locations were chosen for tracking stations: Eastern Europe (Russia, Bear Lakes), Asia (Kazhahstan, Baikonur), Northern Caucasus (Russia, Zelenchyuk) and Siberia (Russia, Ussuriisk).

In addition to the above, it is also necessary to mention the limitations for low orbits associated with the possible overlap of space interferometer field of view (FOV) by the Earth. Table \ref{tab:angs} shows the overlapping angles for the given orbit radius. This angle is estimated as $\alpha = \arccos({R_{\bigoplus}/R_{o}})$, where $R_{\bigoplus}$~=~6378~km -- the Earth's radius, $R_{o}$ -- orbit radius.

\begin{table}[h]
    \centering
    \begin{tabular}{|c|c|c|c|c|}
    \hline
    Orbit radius, km & 7500 & 8000 & 22500 & 23000 \\ \hline
    Angle, deg       & 31.7 & 37.1 & 70.9  & 71.4  \\ \hline
    \end{tabular}
    \caption{\label{tab:angs}The angle between source direction and tangent to the Earth surface ($R_{\bigoplus} \cong 6378$ km).}
\end{table}

\subsection{Synthetic Simulations}
\begin{figure*}
    \centering
    \includegraphics[width=1\linewidth]{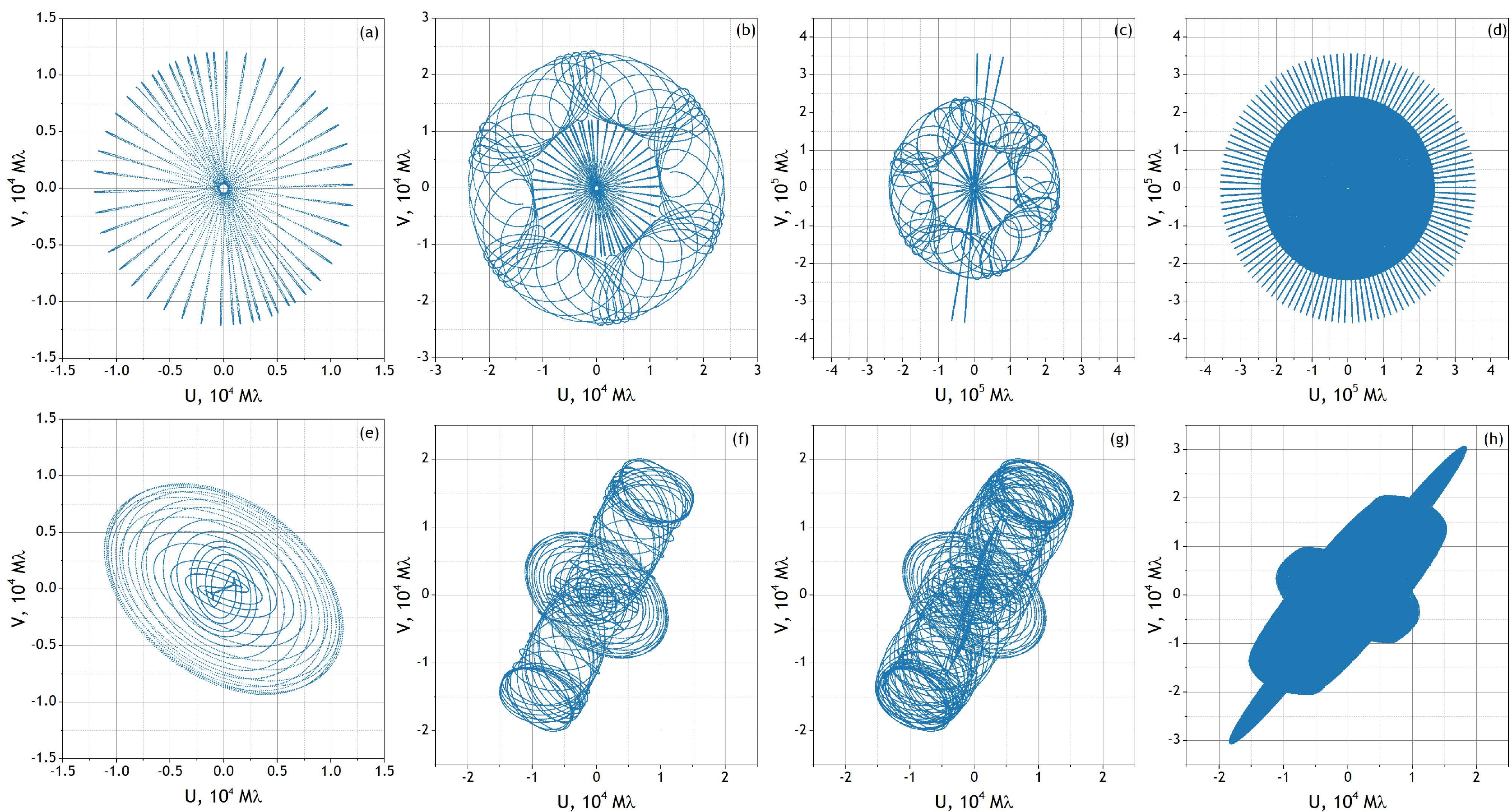}
    \caption{(u,v) coverage for: Sgr~A* -- (a) -- two satellites (24 hours); (b) -- three satellites (24 hours); (c) -- four satellites (24 hours); (d) -- four satellites (full span, 12 days); M87 -- (e) -- two satellites (24 hours); (f) -- three satellites (24 hours); (g) -- four satellites (24 hours); (h) -- four satellites (full span, 12 days). (u,v) time sampling is 20 s.}
    \label{fig:coverage}
\end{figure*}

\begin{figure*}
    \centering
    \includegraphics[width=0.7\linewidth]{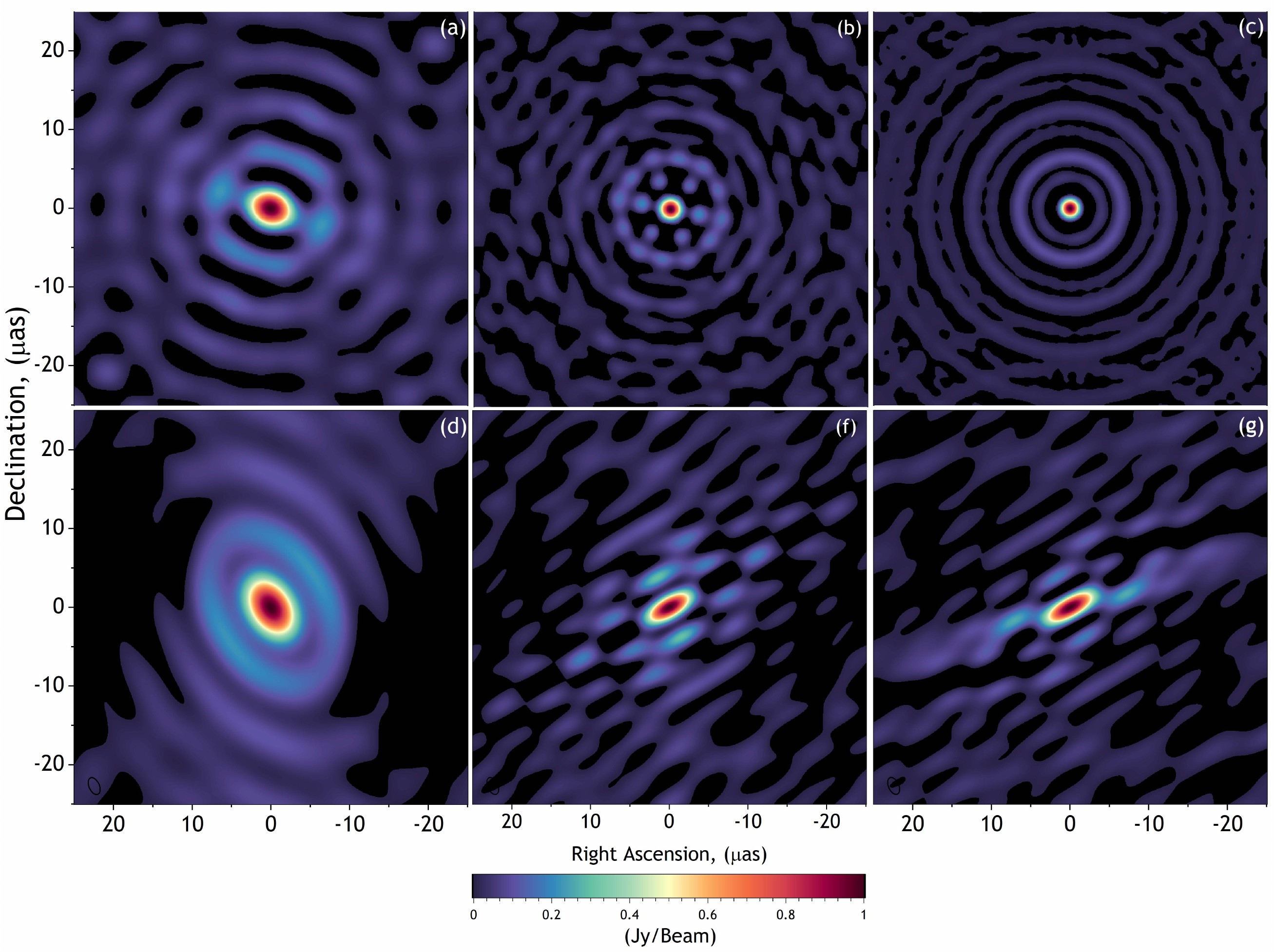}
    \caption{Synthesized beam for (u,v) coverages displayed in Fig.~\ref{fig:coverage} 690~GHz: Sgr A* -- (a) -- two satellites, beam size 4.8$\times$4.0~$\mu$as; (b) -- three satellites, beam size 2.3$\times$2.3~$\mu$as; (c) -- four satellites, beam size 2.0$\times$2.0~$\mu$as.; M87 -- (d) -- two satellites, beam size 7.1$\times$4.8~$\mu$as; (e) -- three satellites, beam size 5.6$\times$2.4~$\mu$as; (f) -- four satellites, beam size 6.1$\times$2.4~$\mu$as.}
    \label{fig:beams}
\end{figure*}

To estimate the performance of the interferometer, its angular resolution and imaging capabilities we have simulated synthetic VLBI observations at 690~GHz of the Sgr~A* source using the obtained orbits. It was done using the Astro Space Locator (ASL) software package\footnote{Latest version of Astro Space Locator software package: https://millimetron.ru/en/for-scientists/astro-space-locator} \cite{Likhachev2020}. At the first stage we calculated $(u,v)$~coverages with 20 s sampling. Fig.~\ref{fig:coverage} shows the corresponding $(u,v)$ for two (a), three (b) and four (c) space telescopes respectively. Full-span $(u,v)$~coverage for two and three telescopes is achieved within $\approx$~20~hours. In the latter case of four telescopes, reaching full-span coverage takes longer (12 days), which is associated with a decrease in the rate of baseline projection evolution for baselines between telescopes located in higher orbits. Fig. \ref{fig:beams} shows synthesized beams for calculated $(u,v)$ coverage: two (a), three (b), and four (c) space telescopes, respectively.

Speaking about the possibilities of observing other sources, we provide $(u,v)$ coverages (see Fig.~\ref{fig:coverage} (e)-(h)) and synthesized beams (see Fig.~\ref{fig:beams} (d)-(f)) for M87 for the specified interferometer geometry. Despite the lower quality of coverages, the shape and size of the beam (7.1$\times$4.8~$\mu$as, 5.6$\times$2.4~$\mu$as and 6.1$\times$2.4~$\mu$as for two, three and four satellites correspondingly in Fig.~\ref{fig:beams} (d)-(f)), it is still possible to carry out imaging and M87 too.

Then Sgr~A* visibility models were applied \cite{Chernov2021}. Magnetohydrodynamic (MHD) modeling was carried out using the free HARM code \cite{Gammie2003,Noble2006}. We started the initial simulation with an equilibrium hydrostatic torus around the black hole. For our goals we use initial torus with the following geometrical and physical parameters: inner radius of the disk $r_{in} = 6$, radius of maximum pressure $r_{max}=12$, adiabatic constant $\Gamma=4/3$, black hole rotation parameter, spin, $a=0.6$. One loop of the magnetic field was considered, which was set using a vector potential $A_\phi\sim\max(\rho-0.2,0)$, where $\rho$ -- the plasma density. The amplitude of the vector potential was normalized by the plasma parameter, which is equal to $\beta=50$. The simulation was performed from $t=0$ to $t=18150\cdot GM/c^3$, where $G$ -- the gravity constant, $c$ -- the speed of light. Starting at the time of $t\approx1000\cdot GM/c^3$, the solution came to a quasi-stationary solution. The remaining time interval corresponds to a duration of four days for a mass equal to $M\approx4\times10^6 M_\odot$, where $M_\odot$ -- is the Sun's mass, which corresponds to the mass of a black hole in Sgr~A*. The distance to the black hole is  $D=8.2kpc$. The inclination angle approximately is $i=66^\circ$. Taking into account the variability of Sgr~A*, we divided every day into 390 models with approximately equal time intervals of $\approx$221~seconds, i.e. source image was updated every 221~seconds for visibility calculation. \cite{Moscibrodzka2014}. Thus, in our calculations, we use 1560 models covering four days of observation.

For each MHD model, source images were built using the \texttt{grtrans} code \cite{Dexter2016}. Images were constructed at 690 GHz to avoid scattering effects. At this frequency, the flux was normalized to 3~Jy \cite{Dexter2014}. It was assumed that thermal electrons emit synchrotron radiation with a temperature equal to one third of protons' temperature, $T_e=T_p/3$. In the Fig.~\ref{fig:image}, the first row shows examples of images obtained by averaging over 390 models on the first, second, third and fourth day of observation. More information about the MHD model and radiation model construction can be found in \cite{Chernov2021}.

Each of the model maps was applied to the calculated $(u,v)$ coverage. In order to do this, each "pixel" of the map was considered as a delta function with corresponding coordinates and flux in Jy. A set of these delta functions was then transferred onto $(u,v)$ coverage using a discrete Fourier transform (DFT). This produces synthetic VLBI data as a complex visibility function.

We took the field of view of 50$\times$50~$\mu$as for the synthesized beam, 100$\times$100~$\mu$as for the source synthesized image and used natural gridding. In the case of two telescopes at 690~GHz, an achieved angular resolution is 4.8$\times$4.0~$\mu$as (Fig. \ref{fig:beams}(a)), for three telescopes the resolution will be 2.3$\times$2.3~$\mu$as (Fig. \ref{fig:beams}(b)), and for four 2.0$\times$2.0~$\mu$as (Fig. \ref{fig:beams}(c)). After that, the standard CLEAN procedure was performed until the first negative component was found.

\section{Results}
Orbits with an optimal parameter ratio were found. This makes it possible to obtain a relatively symmetrical and high-quality filling of the $(u, v)$ plane within a relatively short period of time of 20~hours. Integrating the orbits in a full force field for ten years showed stable configurations. At the same time, the orbits precess, which makes it possible to observe various sources during the mission lifetime. 

We have estimated spacecraft radio visibility from ground tracking stations. One ground tracking station will observe space radio telescopes for $\approx$5~hours/day for 7500-8000 km orbits and for $\approx$12~hours/day for 22500-23000 km orbits (see Fig.~\ref{fig:vis} (left)). At least two ground tracking stations will observe space radio telescopes for $\approx$2 hours/day for 7500-8000 km orbits and for $\approx$8~hours/day for 22500-23000 km orbits (see Fig.~\ref{fig:vis}). 

We have formulated the requirements for orbit determination to perform successful VLBI data correlation. It is possible to implement the requirements using GNSS satellites and VLBI locations. Having an accuracy of orbit determination of about 153~m in distance and 0.9 cm/s in velocity is acceptable for correlation with 4096~spectral channels, $T_{int}$=100~s and $IF=4$~GHz at 690~GHz.

\begin{figure*}
    \centering
    \includegraphics[width=1\linewidth]{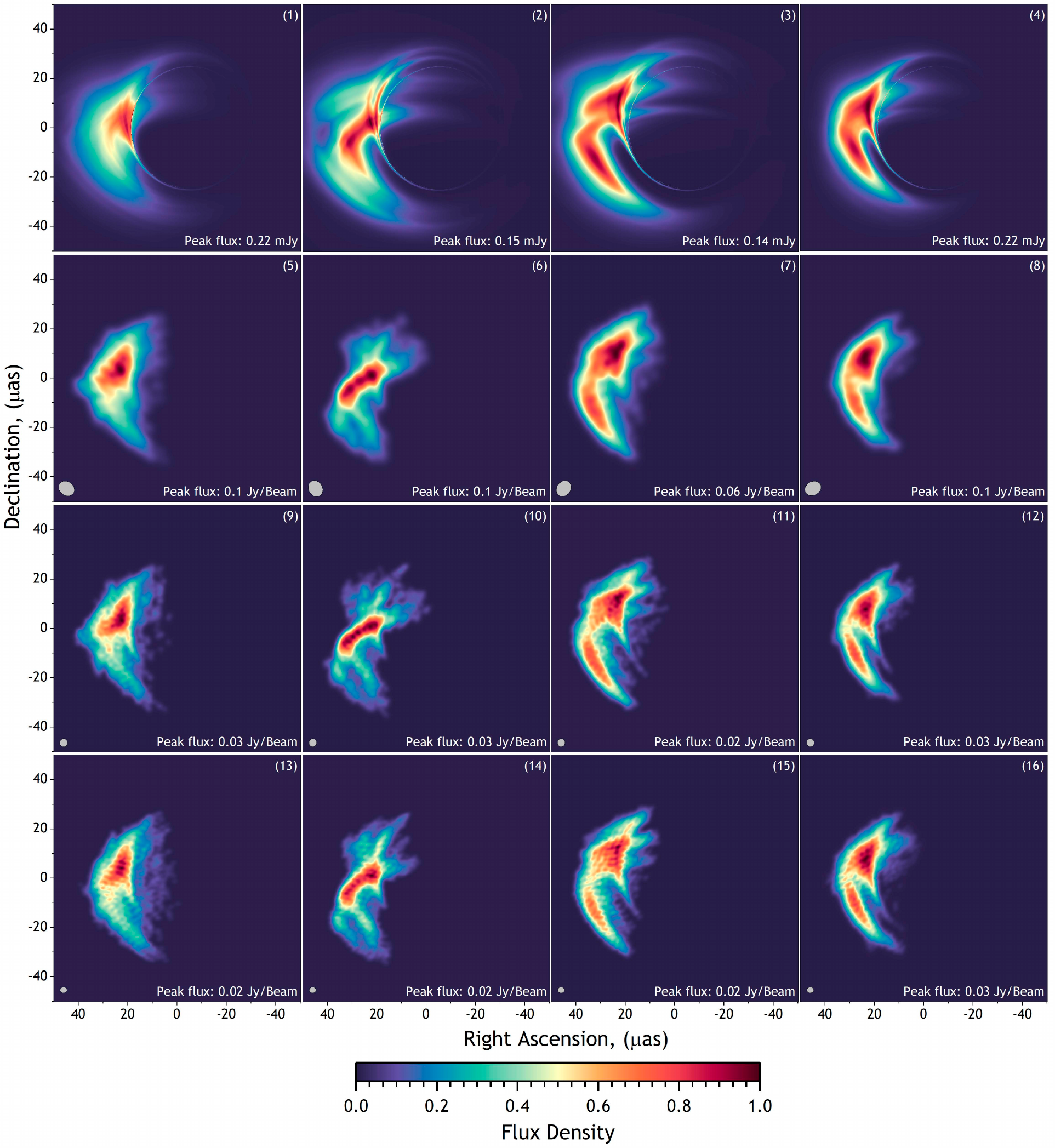}
    \caption{Synthesized image at 690~GHz: (1)-(4) -- models for the 1st, 2nd, 3rd and 4th day of observations; (5)-(8) -- synthesized images for 2 satellites for 1st, 2nd, 3rd and 4th day of observations; (9)-(12) -- synthesized images for 3 satellites for 1st, 2nd, 3rd and 4th day of observations; (13)-(16) -- synthesized images for 2 satellites for 1st, 2nd, 3rd and 4th day of observations. Grey circle shows the size of synthesized beam. Color scale is normalized to peak value of each image.}
    \label{fig:image}
\end{figure*}

\begin{figure}
    \centering
    \includegraphics[width=0.8\linewidth]{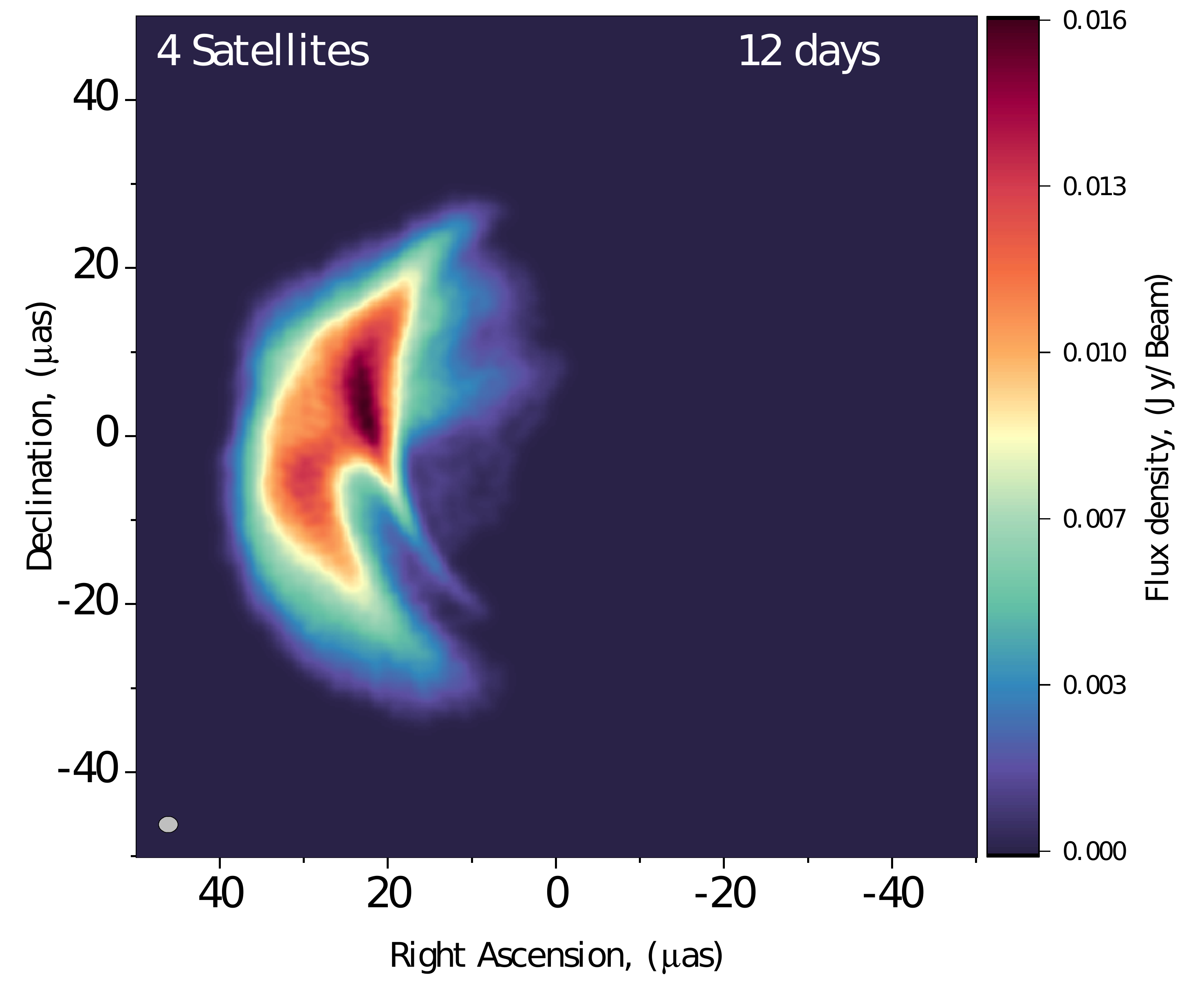}
    \caption{Synthesized image of Sgr~A* at 690~GHz obtained by 4 satellites within 12 days.}
    \label{fig:image12days}
\end{figure}

The results of synthetic image simulations are shown in Fig.~\ref{fig:image} and Fig.~\ref{fig:image12days}. The first row in Fig.~\ref{fig:image} corresponds to the models used in these simulations. From top to bottom, three rows correspond to the configuration of the space interferometer, which in turn corresponds to two, three, and four satellites. From left to right, each column corresponds to its next observation day.

In such configurations, it has been demonstrated that an angular resolution of 4.8$\times$4, 2.3$\times$2.3, 2$\times$2~$\mu$as is achieved for two, three and four space telescopes respectively using natural gridding. The RMS between the obtained images and models in all cases is no more than $1.3\cdot10^{-5}$~Jy, which is an acceptable result of image reconstruction based on the quality of the $(u,v)$~coverage.

In the case when the image is observed during 12 days using 4 satellites, it can be seen that it becomes blurred compared to the images obtained within 24 hours (see Fig.~\ref{fig:image} (13)-(16)). Simulations show the ability to obtain high-quality images of SMBHs in dynamics, particularly in the case of Sgr~A*.

Indeed, the rate of filling of the $(u,v)$ plane and the variability of Sgr~A* do not allow us to speak of a full-fledged observation of the intrinsic variability of this object. Blurring of details is observed according to the accumulation time of a full-span of $(u,v)$ coverage (see Fig.~\ref{fig:image}) and especially for 12 days of observations (see Fig.~\ref{fig:image12days}). However, the proposed geometry at the moment allows you to get a complete span of $(u,v)$ coverage faster than all previously proposed concepts. And here, Sgr~A* was chosen solely in terms of the most challenging object and as a demonstrator, inspired by the studies from \cite{Roelofs2019}. Considering the variability of other objects, M87 may be one of the most promising. Its variability is $7\times~t_{g}$ or $\approx$1000 times longer than for Sgr~A*.

A different initial orientation of the normal to the orbit plane can be chosen. It is possible to consider the direction to the M87 and the expected $(u,v)$ coverage will be the same as for the Sgr~A* simulations presented in this work. In this case, the initial inclination and RAAN for the proposed orbits should be changed to: $i=77.61^{\circ}$, $RAAN=-82.29^{\circ}$ (regular orbits) and $i=-102.39^{\circ}$, $RAAN=-82.29^{\circ}$ (retrograde orbits). Either the direction of the normal to the orbital plane could be selected similar to what is proposed, for example, in EHI \cite{Kudriashov2021a}. To get such configuration, the inclination and RAAN should be as follows: $i=98.31^{\circ}$, $RAAN=-43.0^{\circ}$ (regular orbits) and $i=-81.69^{\circ}$, $RAAN=-43.0^{\circ}$ (retrograde orbits). This will give approximately the same $(u,v)$ coverage for both M87 and Sgr~A*.

\section{Conclusions}
The Event Horizon Imager remains the leader in existing concepts in terms of $(u,v)$~coverage quality. The SMVA concept offered a solution providing one of the best angular resolution at EHT frequencies and coverage in $\approx$24~hours, but with loss of its quality and significant gaps in the $(u,v)$~coverage. CAPELLA is conceptually close to a pure space VLBI system.

The alternative compromise we propose has four spacecraft in circular near-Earth orbits with a radius of 7500-8000~km and 22500-23000~km. The most significant point is that two of these telescopes will operate in retrograde orbits. They will move in the opposite direction to space telescopes operating in regular orbits leading to an increase in the $(u,v)$ coverage filling rate. And the launch of four satellites seems to be the most optimal and feasible option. One launch vehicle will deliver a pair of telescopes into retrograde orbits and the other will deliver a pair into regular orbits. Separately, it should be emphasized that the orbits shown in this paper are stable for at least ten years.

Our synthetic VLBI simulations showed that it would be possible to obtain $(u,v)$~coverage better than in SMVA and close to what is proposed by EHI, but in a significantly shorter time interval $\approx$20~hours versus one month full span (EHI). Moreover, for four satellites, a very effective $(u,v)$~coverage achieved in $\approx$12 days, which is more than twice as fast as EHI's ideal coverage. It is worth noting that although we considered only Sgr~A* in our simulations, the integrated circular orbits precess over time. This means that the mutual orientation of the orbital plane's normal direction with respect to the target source changes. 

All this makes it possible to approach the possibility of observing SMBHs in dynamics. Potentially, it could allow us to obtain even the first detailed images of binary SMBHs systems, images of other SMBHs and sources not been observed before at such frequencies and angular resolution. This makes the concept more universal.

However, it is worth mentioning the constraints of the proposed configuration. The constraints arise from precession (change in RAAN only) and Earth overlapping for R1 will allow observing sources with declination within $\pm$30$^{\circ}$.

Thus, there is an opportunity to further adjust and improve the proposed space interferometer geometry. For example, as indicated in the results, the inclination and RAAN can be changed to make M87 the primary target to $i=77.61^{\circ}$, $RAAN=-82.29^{\circ}$ (normal orbits) and $i=-102.39^{\circ}$ and $RAAN=-82.29^{\circ}$ (retrograde orbits). Following the suggestions of \cite{Kudriashov2021a}, one can image both M87 and Sgr~A* having $i=98.31^{\circ}$, $RAAN=-43.0^{\circ}$ (regular orbits) and $i=-81.69^{\circ}$, $RAAN=-43.0^{\circ}$ (retrograde orbits).

\section*{Data Availability}
The data underlying this article and the results of the simulations will be shared on request to the corresponding author.



\begin{thebibliography}{10}
\expandafter\ifx\csname url\endcsname\relax
  \def\url#1{\texttt{#1}}\fi
\expandafter\ifx\csname urlprefix\endcsname\relax\def\urlprefix{URL }\fi
\expandafter\ifx\csname href\endcsname\relax
  \def\href#1#2{#2} \def\path#1{#1}\fi

\bibitem{Gurvits2020}
L.~I. Gurvits, Space vlbi: from first ideas to operational missions, Advances
  in Space Research 65~(2) (2020) 868--876.

\bibitem{Lovell1999}
J.~E.~J. {Lovell}, H.~{Hirabayashi}, H.~{Kobayashi}, Y.~{Murata}, P.~G.
  {Edwards}, et~al., {Overview and current status of the VSOP mission}, New
  Astronomy Reviews 43~(8-10) (1999) 515--518.

\bibitem{Kardashev2013}
N.~S. {Kardashev}, V.~V. {Khartov}, V.~V. {Abramov}, V.~Y. {Avdeev}, A.~V.
  {Alakoz}, et~al., {``RadioAstron''-A telescope with a size of 300 000 km:
  Main parameters and first observational results}, Astronomy Reports 57 (2013)
  153--194.

\bibitem{Ulvestad2000}
J.~S. Ulvestad, The arise space vlbi mission, Advances in Space Research 26~(4)
  (2000) 735--738.

\bibitem{Mochizuki2003}
N.~{Mochizuki}, H.~{Hirabayashi}, Y.~{Murata}, {Members Of Space Vlbi Wg},
  {VSOP-2; a Next Generation Space-VLBI Mission}, in: Y.~C. {Minh} (Ed.), New
  technologies in VLBI, Vol. 306 of Astronomical Society of the Pacific
  Conference Series, 2003, p. P45.

\bibitem{EHT2019a}
{Event Horizon Telescope Collaboration}, {First M87 Event Horizon Telescope
  Results. I. The Shadow of the Supermassive Black Hole}, Astrophysical Journal
  Letters 875~(1) (2019) L1.

\bibitem{EHT2022a}
{Event Horizon Telescope Collaboration}, {First Sagittarius A* Event Horizon
  Telescope Results. I. The Shadow of the Supermassive Black Hole in the Center
  of the Milky Way}, The Astrophysical Journal Letters 930~(2) (2022) L12.

\bibitem{Palumbo2018}
D.~{Palumbo}, M.~{Johnson}, S.~{Doeleman}, A.~{Chael}, K.~{Bouman},
  {Next-generation Event Horizon Telescope developments: new stations for
  enhanced imaging}, in: American Astronomical Society Meeting Abstracts \#231,
  Vol. 231 of American Astronomical Society Meeting Abstracts, 2018, p. 347.21.

\bibitem{Raymond2021}
A.~W. {Raymond}, D.~{Palumbo}, S.~N. {Paine}, L.~{Blackburn}, R.~{C{\'o}rdova
  Rosado}, et~al., {Evaluation of New Submillimeter VLBI Sites for the Event
  Horizon Telescope}, The Astrophysical Journal Supplement 253~(1) (2021) 5.

\bibitem{Novikov2021}
I.~D. {Novikov}, S.~F. {Likhachev}, Y.~A. {Shchekinov}, A.~S. {Andrianov},
  A.~M. {Baryshev}, et~al., {Objectives of the Millimetron Space Observatory
  science program and technical capabilities of its realization}, Physics
  Uspekhi 64~(4) (2021) 386--419.

\bibitem{Roelofs2019}
F.~{Roelofs}, H.~{Falcke}, C.~{Brinkerink}, M.~{Mo{\'s}cibrodzka}, L.~I.
  {Gurvits}, M.~{Martin-Neira}, V.~{Kudriashov}, M.~{Klein-Wolt}, R.~{Tilanus},
  M.~{Kramer}, L.~{Rezzolla}, {Simulations of imaging the event horizon of
  Sagittarius A* from space}, Astronomy and Astrophysics 625 (2019) A124.

\bibitem{Hong2014}
X.~Hong, Z.~Shen, T.~An, Q.~Liu, The chinese space millimeter-wavelength vlbi
  array—a step toward imaging the most compact astronomical objects, Acta
  Astronautica 102 (2014) 217--225.

\bibitem{Fish2020}
V.~L. Fish, M.~Shea, K.~Akiyama, Imaging black holes and jets with a vlbi array
  including multiple space-based telescopes, Advances in Space Research 65~(2)
  (2020) 821--830.

\bibitem{Kudriashov2021a}
V.~{Kudriashov}, M.~{Martin-Neira}, F.~{Roelofs}, H.~{Falcke}, C.~{Brinkerink},
  et~al., {An Event Horizon Imager (EHI) Mission Concept Utilizing Medium Earth
  Orbit Sub-mm Interferometry}, Chinese Journal of Space Science 41~(2) (2021)
  211--233.

\bibitem{Kudriashov2021b}
V.~{Kudriashov}, M.~{Martin-Neira}, I.~{Barat}, P.~M. {Iglesias},
  E.~{Daganzo-Eusebio}, et~al., {System Design for the Event Horizon Imaging
  Experiment Using the PECMEO Concept}, arXiv e-prints (2021) arXiv:2105.06901.

\bibitem{Gurvits2021}
L.~I. {Gurvits}, Z.~{Paragi}, V.~{Casasola}, J.~{Conway}, J.~{Davelaar},
  et~al., {THEZA: TeraHertz Exploration and Zooming-in for Astrophysics},
  Experimental Astronomy 51~(3) (2021) 559--594.

\bibitem{Gurvits2022}
L.~I. Gurvits, Z.~Paragi, R.~I. Amils, I.~{van Bemmel}, P.~Boven, et~al., The
  science case and challenges of space-borne sub-millimeter interferometry,
  Acta Astronautica 196 (2022) 314--333.

\bibitem{Kurczynski2022}
P.~{Kurczynski}, M.~D. {Johnson}, S.~S. {Doeleman}, K.~{Haworth}, E.~{Peretz},
  et~al., {The Event Horizon Explorer mission concept}, in: L.~E. {Coyle},
  S.~{Matsuura}, M.~D. {Perrin} (Eds.), Space Telescopes and Instrumentation
  2022: Optical, Infrared, and Millimeter Wave, Vol. 12180 of Society of
  Photo-Optical Instrumentation Engineers (SPIE) Conference Series, 2022, p.
  121800M.
\newblock \href {https://doi.org/10.1117/12.2630313}
  {\path{doi:10.1117/12.2630313}}.

\bibitem{Trippe2023}
S.~Trippe, T.~Jung, J.-W. Lee, W.~Kang, J.-Y. Kim, J.~Park, J.~A. Hodgson,
  Capella: A space-only high-frequency radio vlbi network formed by a
  constellation of small satellites (2023).
\newblock \href {http://arxiv.org/abs/2304.06482} {\path{arXiv:2304.06482}}.

\bibitem{Andrianov2021}
A.~S. {Andrianov}, A.~M. {Baryshev}, H.~{Falcke}, I.~A. {Girin}, T.~{de
  Graauw}, et~al., {Simulations of M87* and Sgr A* imaging with the Millimetron
  Space Observatory on near-Earth orbits}, MNRAS 500~(4) (2021) 4866--4877.
\newblock \href {http://arxiv.org/abs/2006.10120} {\path{arXiv:2006.10120}},
  \href {https://doi.org/10.1093/mnras/staa2709}
  {\path{doi:10.1093/mnras/staa2709}}.

\bibitem{Johnson2023}
M.~D. Johnson, K.~Akiyama, L.~Blackburn, K.~L. Bouman, A.~E. Broderick, et~al.,
  \href{https://www.mdpi.com/2075-4434/11/3/61}{Key science goals for the
  next-generation event horizon telescope}, Galaxies 11~(3) (2023).
\newblock \href {https://doi.org/10.3390/galaxies11030061}
  {\path{doi:10.3390/galaxies11030061}}.
\newline\urlprefix\url{https://www.mdpi.com/2075-4434/11/3/61}

\bibitem{Ricarte2023}
A.~Ricarte, P.~Tiede, R.~Emami, A.~Tamar, P.~Natarajan,
  \href{https://www.mdpi.com/2075-4434/11/1/6}{The ngeht's role in measuring
  supermassive black hole spins}, Galaxies 11~(1) (2023).
\newblock \href {https://doi.org/10.3390/galaxies11010006}
  {\path{doi:10.3390/galaxies11010006}}.
\newline\urlprefix\url{https://www.mdpi.com/2075-4434/11/1/6}

\bibitem{Emami2023}
R.~Emami, P.~Tiede, S.~S. Doeleman, F.~Roelofs, M.~Wielgus, et~al.,
  \href{https://www.mdpi.com/2075-4434/11/1/23}{Tracing hot spot motion in
  sagittarius a* using the next-generation event horizon telescope (ngeht)},
  Galaxies 11~(1) (2023).
\newblock \href {https://doi.org/10.3390/galaxies11010023}
  {\path{doi:10.3390/galaxies11010023}}.
\newline\urlprefix\url{https://www.mdpi.com/2075-4434/11/1/23}

\bibitem{Likhachev2022}
S.~F. {Likhachev}, A.~G. {Rudnitskiy}, M.~A. {Shchurov}, A.~S. {Andrianov},
  A.~M. {Baryshev}, et~al., {High-resolution imaging of a black hole shadow
  with Millimetron orbit around lagrange point l2}, MNRAS 511~(1) (2022)
  668--682.
\newblock \href {http://arxiv.org/abs/2108.03077} {\path{arXiv:2108.03077}},
  \href {https://doi.org/10.1093/mnras/stac079}
  {\path{doi:10.1093/mnras/stac079}}.

\bibitem{Han2012}
S.~T. {Han}, J.~W. {Lee}, J.~{Kang}, C.~S. {Oh}, D.~Y. {Byun}, et~al., {Korean
  VLBI Network receiver optics for simultaneous multi-frequency observation},
  in: Proceedings of the 11th European VLBI Network Symposium \& Users Meeting.
  9-12 October, 2012, p.~59.

\bibitem{Han2017}
S.-T. {Han}, J.-W. {Lee}, B.~{Lee}, M.-H. {Chung}, S.-M. {Lee}, et~al., {A
  Millimeter-Wave Quasi-Optical Circuit for Compact Triple-Band Receiving
  System}, Journal of Infrared 38~(12) (2017) 1487--1501.

\bibitem{Rioja2023}
M.~J. Rioja, R.~Dodson, Y.~Asaki, The transformational power of frequency phase
  transfer methods for ngeht, Galaxies 11~(1) (2023).

\bibitem{Golubev2020}
E.~S. {Golubev}, E.~K. {Kotsur}, M.~Y. {Arkhipov}, A.~V. {Smirnov}, A.~O.
  {Lyakhovec}, et~al., {Primary mirror panels of the Millimetron Space
  Observatory}, in: Society of Photo-Optical Instrumentation Engineers (SPIE)
  Conference Series, Vol. 11451 of Society of Photo-Optical Instrumentation
  Engineers (SPIE) Conference Series, 2020, p. 114510K.
\newblock \href {https://doi.org/10.1117/12.2562838}
  {\path{doi:10.1117/12.2562838}}.

\bibitem{Folkner2014}
W.~M. {Folkner}, J.~G. {Williams}, D.~H. {Boggs}, P.~{Park}, R.~S.
  Interplanetary Network Progress Report 196 (2014) 1--81.

\bibitem{EGM96}
F.~G. {Lemoine}, S.~C. {Kenyon}, J.~K. {Factor}, R.~G. {Trimmer}, N.~K.
  {Pavlis}, et~al., {The Development of the Joint NASA GSFC and the National
  Imagery and Mapping Agency (NIMA) Geopotential Model EGM96},
  NASA/TP-1998-206861, NASA Goddard Space Flight Center Technical Report (Jul.
  1998).

\bibitem{Ippolito2008}
L.~Ippolito, \href{https://books.google.ru/books?id=hfZnkqSPSeAC}{Satellite
  Communications Systems Engineering: Atmospheric Effects, Satellite Link
  Design and System Performance}, Wireless Communications and Mobile Computing,
  Wiley, 2008.
\newline\urlprefix\url{https://books.google.ru/books?id=hfZnkqSPSeAC}

\bibitem{Warner2009}
T.~Warner, G.~Foody, M.~Nellis,
  \href{https://books.google.ru/books?id=zk5656l9ulkC}{The SAGE Handbook of
  Remote Sensing}, Sage Handbooks, SAGE Publications, 2009.
\newline\urlprefix\url{https://books.google.ru/books?id=zk5656l9ulkC}

\bibitem{Cakaj2009}
S.~Cakaj, Practical horizon plane and communication duration for low earth
  orbiting (leo) satellite ground stations, WSEAS Transactions on
  Communications Archive 8 (2009) 373--383.

\bibitem{Li2017}
J.~Li, D.~P. Roy, \href{https://www.mdpi.com/2072-4292/9/9/902}{A global
  analysis of sentinel-2a, sentinel-2b and landsat-8 data revisit intervals and
  implications for terrestrial monitoring}, Remote Sensing 9~(9) (2017).
\newblock \href {https://doi.org/10.3390/rs9090902}
  {\path{doi:10.3390/rs9090902}}.
\newline\urlprefix\url{https://www.mdpi.com/2072-4292/9/9/902}

\bibitem{Andrianov2014}
A.~C. Andrianov, I.~A. Girin, V.~Zharov, V.~Kostenko, Correlator of the fian
  astro space center in radioastron mission, Vestnik NPO imeni S.A. Lavochkina
  (2014) 55--58.

\bibitem{Zhamkov2016}
A.~Zhamkov, V.~Zharov, Improvement of the orbit of the spektr-r spacecraft in
  the radioastron mission on the basis of radio range and doppler measurements,
  Moscow University Physics Bulletin 71 (2016) 299--308.

\bibitem{Likhachev2017}
S.~Likhachev, V.~Kostenko, I.~Girin, A.~Andrianov, V.~Jarov, A.~Rudnitskiy,
  Software correlator for the radioastron mission, Journal of Astronomical
  Instrumentation 6 (06 2017).
\newblock \href {https://doi.org/10.1142/S2251171717500040}
  {\path{doi:10.1142/S2251171717500040}}.

\bibitem{MEO2}
T.~Geng, X.~Su, Q.~Zhao, Meo and heo satellites orbit determination based on
  gnss onboard receiver, Lecture Notes in Electrical Engineering 160 (2012)
  223--234.

\bibitem{MEO3}
H.~Liu, X.~Cheng, G.~Tang, J.~Peng, Gnss performance research for meo, geo, and
  heo, in: J.~Sun, J.~Liu, Y.~Yang, S.~Fan, W.~Yu (Eds.), China Satellite
  Navigation Conference (CSNC) 2017 Proceedings: Volume III, Springer
  Singapore, Singapore, 2017, pp. 37--45.

\bibitem{MEO4}
M.~Guan, T.~Xu, M.~Li, F.~Gao, D.~Mu, Navigation in geo, heo, and lunar
  trajectory using multi-gnss sidelobe signals, Remote Sensing 14~(2) (2022).

\bibitem{Duev2012}
D.~Duev, G.~Calves, S.~Pogrebenko, L.~Gurvits, G.~Cimo, T.~Bocanegra-Bahamón,
  Spacecraft vlbi and doppler tracking: Algorithms and implementation,
  Statistica Neerlandica - STAT NEERL 541 (03 2012).
\newblock \href {https://doi.org/10.1051/0004-6361/201218885}
  {\path{doi:10.1051/0004-6361/201218885}}.

\bibitem{Klopoter2020}
G.~Kłopotek, T.~Hobiger, R.~Haas, T.~Otsubo, Geodetic vlbi for precise orbit
  determination of earth satellites: a simulation study, Journal of Geodesy 94
  (2020) 56--.
\newblock \href {https://doi.org/10.1007/s00190-020-01381-9}
  {\path{doi:10.1007/s00190-020-01381-9}}.

\bibitem{Likhachev2020}
S.~Likhachev, I.~Girin, V.~Y. Avdeev, A.~Andrianov, M.~Andrianov, et~al., Astro
  space locator — a software package for vlbi data processing and reduction,
  Astronomy and Computing 33 (2020) 100426.
\newblock \href {https://doi.org/https://doi.org/10.1016/j.ascom.2020.100426}
  {\path{doi:https://doi.org/10.1016/j.ascom.2020.100426}}.

\bibitem{Chernov2021}
S.~V. {Chernov}, {Restrictions on the Parameters of Black Hole and Plasma in
  the Vicinity of the Source Sagittarius A*}, Astronomy Reports 65 (2021)
  110--125.
\newblock \href {https://doi.org/10.1134/S1063772921020013}
  {\path{doi:10.1134/S1063772921020013}}.

\bibitem{Gammie2003}
G.~F. Gammie, J.~C. McKinney, G.~Tóth, {HARM: A Numerical Scheme for General
  Relativistic Magnetohydrodynamics }, The Astrophysical Journal 589 (2003)
  444--457.
\newblock \href {https://doi.org/10.1086/374594} {\path{doi:10.1086/374594}}.

\bibitem{Noble2006}
S.~C. Noble, C.~F. Gammie, J.~C. McKinney, L.~Del~Zanna, {Primitive Variable
  Solvers for Conservative General Relativistic Magnetohydrodynamics }, The
  Astrophysical Journal 641 (2006) 626--637.
\newblock \href {https://doi.org/10.1086/500349} {\path{doi:10.1086/500349}}.

\bibitem{Moscibrodzka2014}
M.~{Mo{\'s}cibrodzka}, H.~{Falcke}, H.~{Shiokawa}, C.~F. {Gammie},
  {Observational appearance of inefficient accretion flows and jets in 3D GRMHD
  simulations: Application to Sagittarius A*}, Astronomy and Astrophysics 570
  (2014) A7.
\newblock \href {http://arxiv.org/abs/1408.4743} {\path{arXiv:1408.4743}},
  \href {https://doi.org/10.1051/0004-6361/201424358}
  {\path{doi:10.1051/0004-6361/201424358}}.

\bibitem{Dexter2016}
J.~Dexter, {A public code for general relativistic, polarised radiative
  transfer around spinning black holes}, MNRAS 462 (2016) 115--136.
\newblock \href {https://doi.org/10.1093/mnras/stw1526}
  {\path{doi:10.1093/mnras/stw1526}}.

\bibitem{Dexter2014}
J.~Dexter, B.~Kelly, G.~C. Bower, D.~P. Marrone, J.~Stone, R.~Plambeck, {An 8 h
  characteristic time-scale in submillimetre light curves of Sagittarius A*},
  MNRAS 442 (2014) 2797--2808.
\newblock \href {https://doi.org/10.1093/mnras/stu1039}
  {\path{doi:10.1093/mnras/stu1039}}.

\end{thebibliography}
\end{document}